\documentclass[aps,prb,preprint,showpacs]{revtex4-2}
\usepackage[utf8x]{inputenc}
\usepackage[english]{babel}
\usepackage[dvips]{graphicx}
\usepackage{url}
\usepackage{bm}
\usepackage{amsmath,amssymb}
\usepackage{lineno}

\usepackage{color,soul}
\setpagewiselinenumbers

\usepackage[unicode]{hyperref}

\hypersetup{pdftitle=Is orbital ordering in crystals a magnetic phase transition?,
pdfstartview=FitV,
citecolor=magenta,
colorlinks=true,
linkcolor=blue}

\newcommand{\vc}{\mathbf{c}}

\newcommand{\ve}{\mathbf{e}}

\newcommand{\vj}{\mathbf{j}}

\newcommand{\vl}{\mathbf{l}}
\newcommand{\vm}{\mathbf{m}}

\newcommand{\vvr}{\mathbf{r}}

\newcommand{\vcj}{\mathbf{J}}
\newcommand{\vcm}{\mathbf{M}}

\newcommand{\cu}{\mathrm{Cu}}
\newcommand{\chr}{\mathrm{Cr}}
\newcommand{\fe}{\mathrm{Fe}}
\newcommand{\nc}{\mathrm{Ni}}
\newcommand{\ox}{\mathrm{O}}

\begin{document}

\title{Is orbital ordering in crystals a magnetic phase transition?}

\author{Borlakov Kh. Sh.}
\affiliation{Department of Physics, 
North Caucasian State Academy, 
Cherkessk, 369000, Russia,
Stavropolskaya st., 36, e-mail:borlakov@mail.ru}
\author{Ediev D. M.}
\affiliation{Department of Mathematics, 
North Caucasian State Academy, 
Cherkessk, 369000, Russia,
Stavropolskaya st., 36}
\author{Borlakova A. Kh.}
\affiliation{Department of Informatics, 
North Caucasian State Academy, 
Cherkessk, 369000, Russia,
Stavropolskaya st., 36}

\date{\today}

\maketitle

\begin{center}
  \Large \textbf{Spontaneous time-reversal symmetry breaking orbital order}
\end{center}

\textbf{abstract.}
It is shown that the time-reversal symmetry is broken during the phase
transition causing orbital ordering in a crystal with Jahn-Teller (J-T)
ions.
This means that such phase transitions are magnetic ones.
As an example we have considered the spinel crystals containing $T_{2g}$ ions:
$\nc\chr_2\ox_4$ ($T_{orb}=310$ K, $T_c=70$ K) and $\cu \chr_2 \ox_4$
($T_{orb}=860$ K, $T_c=135$ K), where the orbital ordering precedes the
spin ordering, which occurs at lower temperatures.
The spin-ordering transition is naturally included into our general
group-theoretical scheme of the phase transition.
Based on the symmetry group of the parent phase, the symmetry groups for
possible orbital- and spin-ordered phases are found.
We present arguments in favor of the idea that the orbital ordering occurs due
to the superexchange between J-T ions, and the spin ordering is caused by the
spin-orbital interaction.
At the same time, the displacements of atoms of the coordination polyhedron do
not lead to the removal of the orbital degeneracy, and they are merely an
accompanying effect.

\hrulefill


\section{Introduction}
\label{sec:introduction}

Break-up of the crystal symmetry under the time-reversal means that phases
with lower symmetry, which emerge from the parent phase of the crystal during
the phase transition, are magnetically ordered.
In this article we present arguments in favor of the idea that the orbital
ordering in crystals with J-T ions is a magnetic phase transition.
Our treatment is based on the group-theoretical analysis performed within the
framework of Landau theory of phase transitions.
The parameters of Landau theory are the mean values of quantum operators or
their classical limits.
For example, the spin is replaced with the vector of magnetic momentum.
However, parameters, describing degenerate orbital states, are essentially
quantum.
These are the wave functions of the $3d$-level of the J-T ion, which have no
classical limits.
Therefore, the way these orbital parameters enter the scheme of Landau theory
is of key importance.
An adequate solution of this problem was proposed in 1989 in the work
\cite{Gurin_jetp89}.
There, the local density matrix $\hat{\rho}$ was introduced, which describes
the mixed state of the J-T ion on a lattice node.
Using $\hat{\rho}$ one can calculate mean values of the charge and current
densities on the node.
The work \cite{Izyumov_pt98} provides more details of the local density matrix
approach.
It also describes the problems, which can be addressed using this approach.
Unfortunately,
the remarkable fact of the occurrence of a non-zero current on the J-T node below
the point of the orbital ordering was not properly recognized by authors of
\cite{Gurin_jetp89,Jirak_prb92},
who did not draw any conclusion out of it.
Most of all, the conclusion on the magnetic nature of the orbital ordering was
not made.
After quite a long time since the publication of the work
\cite{Gurin_jetp89,Izyumov_pt98,Jirak_prb92}
the formalism of the local density matrix has not attracted an attention from the
researchers working on the theory of charge and orbital ordering in J-T
crystals.
This fact was the main motivation of the presented work.
Meanwhile, the occurrence of the orbital current in the orbitally-ordered
phase is the clear evidence that the orbital ordering is the magnetic 
phase transition \cite{Landau_T8}.
As an example application of our theory we consider the spinel crystals
$\nc\chr_2 \ox_4$  ($T_{orb}=310$ K, $T_s=70$ K) and $\cu \chr_2 \ox_4$ ($T_{orb}=860$ K,
$T_s=135$ K).
In those crystals the orbital ordering precedes the spin-ordering, which
happens at a lower temperature.
For temperatures $T_s < T < T_{orb}$ only orbital ordering takes place, while
below $T_s$ the orbital and spin-ordering coexist.
Landau theory of phase transitions allows one to describe the charge-, spin-
and orbital ordering 
within the  unified symmetry model with the subordination scheme
``group-subgroup''. 
Moreover, it makes it possible to identify the interactions,
driving each of the above phase transition.
This question was puzzling many researchers working on the theory of orbital
ordering \cite{Streltsov_2017}.

\section{Local density matrix for degenerate $T_{2g}$-states}
\label{sec:II}

The introduction of the local (nodal) $3 \times 3$ density matrix $\hat\rho$
for J-T nodes is based on the following arguments \cite{Gurin_jetp89,Jirak_prb92}.
In a disordered phase the probability for an electron to be in any of the
three possible states is the same.
Below the ordering point, i.e., for $T<T_{orb}$, the corresponding
probabilities are different.
Thus, it is reasonable to assume that the variation of the density matrix
$\Delta \hat\rho$ is the function of modulus of the critical ordering
parameter (OP) $\eta$, i.e., $\Delta\hat{\rho} =\Delta\hat{\rho}(\eta)$, so
we can write
\begin{equation}
  \label{eq:def-rho}
  \hat\rho = \frac{1}{3} \hat E + \Delta \hat\rho(\eta),
\end{equation}
where $\hat E$ is the unit $3 \times 3$-matrix, the term $\hat E/3$ is the
density matrix of a disordered state, and the second term $\Delta\hat \rho$ is
the traceless part of the density matrix, emerging below the point of the
phase transition.
According to (\ref{eq:def-rho}), the parameter $\Delta\hat{\rho}$ describes
the occurrence of new physical properties of the low-symmetric phase.
Let us decompose $\Delta\hat{\rho}$ over the set of Hermitian traceless
matrices $\hat{\lambda}_a$, which form an irreducible basis in the
eight-dimensional space of such matrices.
Indeed, for a $3\times3$ Hermitian matrix $A_{jk}$ we have that the
three elements on the main diagonal $A_{jj}$ are real and the three
complex elements below the main diagonal are complex conjugated of those above
it: $A_{jk} = A^{*}_{kj}$, $j<k$.
Thus, a $3 \times 3$ Hermitian matrix has 9 independent real parameters minus
one, as demanded by the condition $\sum_{j} A_{jj} =0$.
As a consequence, we can write
\begin{equation}
  \label{eq:rho-exp}
  \Delta\hat{\rho} = \sum_{\alpha=1}^{8} A_\alpha (\eta) \,
  \hat{\lambda}_\alpha,
  \quad
  A_\alpha (\eta) = \langle \hat\lambda_\alpha \rangle
  = \mathrm{Sp}\; \langle \hat{\rho} \, \hat{\lambda}_\alpha \rangle,
\end{equation}
where $\alpha$ is the number of the irreducible representation (IR), entering
the tensor representation over $\Delta\hat{\rho}$.
Coefficients $A_\alpha (\eta)$ depend on the modulus of the critical OP $\eta$,
and they vanish for zero OP: $A_\alpha(0)=0$.
As elements of the matrix basis $\hat{\lambda}_\alpha$ we can choose the
Gell-Mann matrices. 
Their standard form is  \cite{Gurin_jetp89,rumer_unit_sym}
\begin{equation}
  \label{eq:def-lambda}
  \begin{split}
\lambda_1 &= \begin{pmatrix} 0 & 1 & 0 \\ 1 & 0 & 0 \\ 0 & 0 & 0 \end{pmatrix},\;
\lambda_2 = \begin{pmatrix} 0 & -i & 0 \\ i & 0 & 0 \\ 0 & 0 & 0 \end{pmatrix}, \;
\lambda_3 = \begin{pmatrix} 1 & 0 & 0 \\ 0 & -1 & 0 \\ 0 & 0 & 0 \end{pmatrix},\;
\lambda_4 = \begin{pmatrix} 0 & 0 & 1 \\ 0 & 0 & 0 \\ 1 & 0 & 0 \end{pmatrix}, \\
\lambda_5 &= \begin{pmatrix} 0 & 0 & -i \\ 0 & 0 & 0 \\ i & 0 & 0 \end{pmatrix},\;
\lambda_6 = \begin{pmatrix} 0 & 0 & 0 \\ 0 & 0 & 1 \\ 0 & 1 & 0 \end{pmatrix},\;
\lambda_7 = \begin{pmatrix} 0 & 0 & 0 \\ 0 & 0 & -i \\ 0 & i & 0 \end{pmatrix},\;
\lambda_8 = \frac{1}{\sqrt{3}}
  \begin{pmatrix} 1 & 0 & 0 \\ 0 & 1 & 0 \\ 0 & 0 & -2 \end{pmatrix}.
  \end{split}
\end{equation}
Inserting (\ref{eq:def-lambda}) into (\ref{eq:rho-exp}) we obtain the
expression for the variation of the density matrix in terms of the
$A$-parameters depending on the critical OP:
\begin{equation}
  \label{eq:rho-A}
  \Delta\hat{\rho} =
  \begin{pmatrix}
    A_3 + \frac{A_8}{\sqrt 3} & A_1 - i A_2 & A_4 - i A_5 \\
    A_1 + i A_2 & -A_3 + \frac{A_8}{\sqrt 3} & A_6 - i A_7 \\
    A_4+i A_5 & A_6 + i A_7 & - 2 \frac{A_8}{\sqrt 3} 
  \end{pmatrix}.
\end{equation}
The variation of the density matrix (\ref{eq:rho-A}) allows one to calculate
the average values of the quantum-mechanical operators, defining the tensor
properties of the J-T node in an orbitally ordered phase.
Let us first analyze the corresponding quantum-mechanical operators.


\section{Averaged charge and current density on the J-T node}
\label{sec:III}

\begin{figure}[ht]
  \centering
  \includegraphics[width=0.6\textwidth]{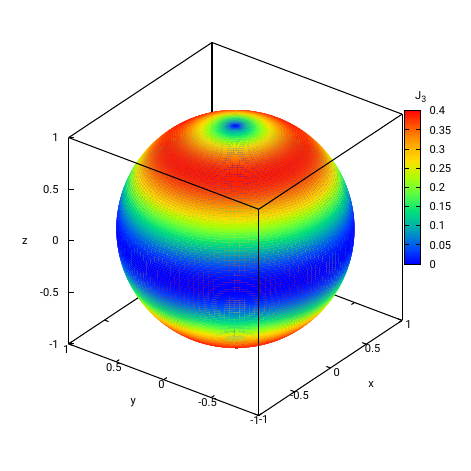}
  \caption{Intensity of the $z$-component of the third basis current
    $\mathbf{J}_3$ on the $t_{2g}$-center (arbitrary units). As $x$- and
    $y$-components of $\mathbf{J}_3$ are equal to zero, the vector
    $\mathbf{J}_3$ is directed along the $z$-axis.
    Maximum intensity corresponds to the polar angles
    $\theta_{max}=\pm 15^{o}16'$.
  }
  \label{fig:J3}
\end{figure}

The main physical phenomena inherent to the orbital ordering are the variation
of the electronic density on the J-T node and the emergence of the orbital
current on the J-T node (Ampere's current).
The variation of the charge density in the orbitally ordered phase is
described by the non-zero mean value of the charge density operator $\hat{d}$,
whose matrix elements are
\begin{equation}
  \label{eq:d_ik}
  d_{ik} e \langle \varphi_i | {\hat d} | \varphi_k \rangle
  = e \, \varphi_i \, \varphi^{*}_k.
\end{equation}
The emergence of the stationary orbital current on the J-T node is described by
non-zero mean value of the current density operator $\mathbf{J}$, whose matrix
elements are \cite{Balcar_1975}:
\begin{equation}
  \label{eq:J_ik}
  J_{ik}= e \langle \varphi_i | \mathbf{\hat J} | \varphi_k \rangle
  =\frac{i e \hbar}{2m} \left(
    \varphi_k \nabla \varphi^{*}_i
   -\varphi_i \nabla \varphi^{*}_k \right).
\end{equation}
In the above formula $e$ and $m$ are the electric charge and mass of the
electron.
Mean value of the charge density operator on the J-T node are calculated using
the formulae
$\langle \hat{d} \rangle = \mathrm{Sp}\, \hat{\rho} \hat{d}$, and analogously
for $\langle \mathbf{J}\rangle$.
 The triplet of basis functions for $T_{2g}$-states has the form:
\begin{equation}
  \label{eq:phi-s}
  \varphi_1 = \sigma \frac{yz}{r^{2}},\;
  \varphi_2 = \sigma \frac{zx}{r^{2}},\;
  \varphi_1 = \sigma \frac{xy}{r^{2}},\;
  \sigma = \sqrt\frac{15}{4 \pi}.
\end{equation}
As is seen, all basis functions are real-valued.
Therefore, the diagonal elements of the matrix of the current density are zero
and its off-diagonal elements are:
\begin{equation}
  \label{eq:J-s}
  \vj_{23} = \frac{i e \hbar}{2m} \left(
    \varphi_2 \nabla \varphi^{*}_3
   -\varphi_3 \nabla \varphi^{*}_2 \right), \;
  \vj_{13} = \frac{i e \hbar}{2m} \left(
    \varphi_3 \nabla \varphi^{*}_1
    -\varphi_1 \nabla \varphi^{*}_3 \right), \;
    \vj_{12} = \frac{i e \hbar}{2m} \left(
    \varphi_2 \nabla \varphi^{*}_1
   -\varphi_1 \nabla \varphi^{*}_2 \right).
\end{equation}
Taking the trace of the product of the density matrix and the matrix of the
current density, we arrive at the expression for the mean
current density on the J-T node:
\begin{equation}
  \label{eq:def-j}
  \vj = A_7 \mathbf{J}_1 - A_5 \mathbf{J}_2 + A_2 \mathbf{J}_3,
\end{equation}
where the basis currents $\mathbf{J}_1, \mathbf{J}_2, \mathbf{J}_3$ are
\begin{equation}
  \label{eq:def-J}
  \mathbf{J}_1 = \vj_{23}, \quad
  \mathbf{J}_2 = \vj_{13}, \quad
  \mathbf{J}_3 = -\vj_{23}.
\end{equation}
Let us analyze the properties of the current streamlines by taking $\mathbf{J}_3$
as an example.
In spherical coordinates  $\mathbf{J}_3$ has only one non-zero component,
which is
\begin{equation}
  \label{eq:J3-phi}
  J_{3\varphi} = \frac{e \hbar}{m} \frac{45}{4 \pi} \sin \theta\,
  \cos^{3} \theta.
\end{equation}
The map of the current intensity for $\mathbf{J}_3$ is shown in Figure~\ref{fig:J3}.
Note that  $\mathbf{J}_3$ reaches its maximum at the angles
$\theta_{max}= \pm \arctan (1/\sqrt 2) \approx \pm 15^o 16'$.
Thus, the main contribution to the magnetic moment stems from the
infinitesimal neighborhood of a two-part cone whose vertex is at the origin of
the coordinate frame and the symmetry axis coincides with the $z$-axis.
The streamlines, corresponding to the maximal current, are wound on the
surface of that cone.
The maxima of the two other basis currents, $\mathbf{J}_1$ and $\mathbf{J}_2$ are
located on the conical surfaces whose axes are the $x$- and $y$-axes
of the coordinate frame, respectively.
Calculations, analogous to the presented above, lead to the following expression
for the mean charge density on the J-T node:
\begin{equation}
  \label{eq:mean-d}
  \langle \hat{d}\rangle = e \biggl[
  \left( \frac{1}{3} + A_3 \right) (\varphi^2_1 - \varphi^2_2)
  +  \left( \frac{1}{3} + \frac{A_8}{\sqrt 3} \right)
  (\varphi^2_1 + \varphi^2_2 - 2 \varphi^2_3)
  + 2 A_1 \varphi_1 \varphi_2 + A_4 \varphi_1 \varphi_3 + 2 A_6 \varphi_2
  \varphi_3
  \biggr].
\end{equation}
It is known, that the triplet of functions (\ref{eq:phi-s}) is transformed
under the IR $T_{2g}$ of the point group $O_h$.
Calculating the action of the generator matrices of IR $T_{2g}$ on the triplet
(\ref{eq:J-s}), one could see that the product $\varphi_i \varphi_j$ transforms
into each other, and matrices of the corresponding transformations coincide
with the matrices of $T_{2g}$.
This leads to the conclusion that the factors at the three last terms in
(\ref{eq:mean-d}) are equal:
$A_1=A_4=A_6$.

Similarly, functions $\varphi_1^2-\varphi_2^2$ and $\varphi_1^2+\varphi^2_2 -
2 \varphi^2_3$ form a doublet, transforming under the IR $E_g$ of the
group $O_h$.
Consequently, $\sqrt 3 A_3=A_8$ and equation (\ref{eq:mean-d}) takes the form
\begin{equation}
  \label{eq:mean-d-a}
  \langle \hat{d}\rangle = e \biggl[
  \left( \frac{1}{3} + A_3 \right) (\varphi^2_1 - \varphi^2_2)
  +  2 A_1 (\varphi_1 \varphi_2 + \varphi_1 \varphi_3 + \varphi_2 \varphi_3)
  \biggr].
\end{equation}
The charge ordering on J-T nodes means that the average value of the electric
charge on the J-T node, which is equal to the electron charge in a disordered
phase, becomes different for two 2(a) Wyckoff positions
in a spinel structure having spatial symmetry of the point group $O_h$.
In the simplest case of the 2(a) position splitting without
the change of the translational symmetry
the charge of one node will be larger than $e$: $q_1=e (1+\varepsilon)$, while the
charge of the other node will be smaller than $e$: $q_2=e (1- \varepsilon)$.

Key elements of the phenomenological theory of phase transitions are
integral quantities characterizing nodes of the crystal lattice,
rather than functions similar to the basis currents (\ref{eq:def-J}).
Therefore, we have to calculate the basis orbital magnetic moments $m_1$,
$m_2$, $m_3$ induced by these currents.


\section{ Basis magnetic moments}
\label{sec:IV}

The mean stationary current $\mathbf{J}(\vvr)$, localized on the J-T ion,
defines a magnetization $\mathbf{M}(\vvr)$.
The integral of $\mathbf{M}(\vvr)$ over the volume of the atom is the magnetic
momentum $\vm$.
Vectors $\mathbf{J}(\vvr)$ and $\mathbf{M}(\vvr)$ are connected with each
other by means of the differential equation \cite{Landau_T3}:
\begin{equation}
  \label{eq:MJ}
  \nabla \times \mathbf{M} (\vvr)= \mathbf{J}(\vvr).
\end{equation}
This equation follows from the system of macroscopic Maxwell equations and its
solution can be written as \cite{Landau_T3}:
\begin{equation}
  \label{eq:M-via-J}
  \mathbf{M}(\vvr) = \vvr \times \mathbf{J}(\vvr) + \nabla g(\vvr),
\end{equation}
where $g(\vvr)$ is an arbitrary function defined by a gauge convention.
The gauge convention corresponding for the J-T ion in a crystal is given in
the Appendix~A.
Calculating the volume integral of the equation (\ref{eq:M-via-J}), we obtain
an expression for the orbital magnetic momentum  $\vm$:
\begin{equation}
  \label{eq:m-int}
  \vm = \int \mathbf{M}(\vvr) \, d \vvr 
= \int \vvr \times \mathbf{J}(\vvr) \, d \vvr.
\end{equation}
It is shown in Appendix~\ref{sec:m-moments-gauge-inv}, that the Trammel gauge
demands the integration volume to be finite.
According to equation (\ref{eq:r-x}), the cutoff radius is 
$r_1 \approx 3.4 a_0$.
The explicit dependence of the basis currents $\mathbf{J}_i$ on coordinates
$x,y,z$ can be determined by 
inserting expressions (\ref{eq:phi-s}) for the basis functions
$\varphi_i$ into equations~(\ref{eq:def-J}) for the basis currents. 
Substituting the ensuing expressions into (\ref{eq:m-int}) we can obtain the
numerical values of the magnitudes of the basis magnetic moments $m_1$, $m_2$,
$m_3$.
Details of the computations are given in Appendix~A and the final result is
\begin{equation}
  \label{eq:m-expl}
  \mathbf{m}_i = \frac{e \hbar}{m}\, \tilde{\ve}_i, \quad i=1,2,3,
\end{equation}
where $\tilde{\ve_1}$, $\tilde{\ve_2}$, $\tilde{\ve}_3$ are the unit axial
Cartesian basis vectors.
Note that vectors $\tilde{\ve}_i$ are time-odd, i.e., they change sign under
the time inversion: $\tilde{\ve}_i \to - \tilde{\ve}_i$ for $t \to -t$.

If we employ the vector $\vj(\vvr)$ defined by (\ref{eq:def-j}) as $\mathbf{J}$
in (\ref{eq:m-int}) then, by taking the cross-product of $\vvr$ with the
vector equation (\ref{eq:def-j}) and integrating it over the volume, we arrive
at the expansion of the orbital magnetic moment of the J-T node, similar to
(\ref{eq:def-j}),
\begin{equation}
  \label{eq:m-A}
  \mathbf{m} = A_7 \mathbf{m}_1 -A_5 \mathbf{m}_2 + A_2 \mathbf{m}_3.
\end{equation}
This equation defines the expansion of the tensor characteristic (the magnetic
moment $\mathbf{m}(\vvr)$) of the J-T node over the basis of functions having
the same physical nature.
At first sight it seems that the triplet $(A_7, A_5, A_2)$ could serve as the
three-dimensional OP in the theory of the orbital ordering, as it was proposed
in \cite{Gurin_jetp89,Jirak_prb92}.
However, the true OP, $c$, and the triplet $(A_7, A_5, A_2)$ have different
parity with respect to the time inversion.
Namely, $c$ changes its sign under the time inversion, while the triplet $(A_7,
A_5, A_2)$ from (\ref{eq:m-A}), does not.
Nevertheless, both these three-component quantities behave equally under the
proper rotations: orientations of the OP $c$ and the triplet $(A_7, A_5, A_2)$
are mutually consistent.
Therefore, one can use the symmetry of the OP $c$  to determine the 
symmetry of the triplet $(A_7, A_5, A_2)$ and to simplify expressions for the
corresponding density matrix. 
This is necessary to analyze the level splitting of the $3d$-ion under the
crystal symmetry change.


\section{Symmetry group of a paramagnetic phase and the choice of the critical
  IR, inducing orbital ordering}
\label{sec:V}

So far, we have not used any information on the structure of particular
crystals, whose thermodynamic properties under the phase transitions we were
going to analyze.
The only fact we have implicitly utilized was the result of the crystal field
theory about the splitting of the 5-fold degenerate $3d$-level of a free atom
in the octahedral and tetrahedral fields of the coordination polyhedron.
For future consideration we need to know more details on the crystal structure
of spinels and the representations of the space group $O_h^{7}$, which
describes the symmetry of chromite spinels $\nc\chr_2\ox_4$ and $\cu \chr_2
\ox_4$ in the orbitally disordered phase.
J-T ions $\chr^{2+}$ and $\cu^{2+}$ occupy two most symmetric positions in the
tetrahedral sublattice of the spinel structure.
Coordinates of the Wyckoff's positions 2(a) are $1(000)$, $2(1/4,1/4,1/4)$ in
parts of the edge of the \textbf{fcc} (face-centered cubic) cell.
The volume of the primitive cell is preserved during the orbital
ordering, which means that the orbitally ordered phases are induced by IR's of
the wave vector $k=k_{11}=0$.
Below we use Kovalev's \cite{Kovalev_1993} nomenclature for the wave vectors and IR's.
In the orbitally ordered phase the crystal has tetragonal structure 
\cite{Lenglet_1986,Kino_jpsj66},
and its symmetry group is $D_{4h}^{19}=I4_1/amd$
\cite{Suchomel_prb12}.

Due to the magnetic nature of the orbital ordering, proposed by us, the
initial phase should be considered as paramagnetic, and its symmetry to be
that of the gray paramagnetic Shubnikov group
$O_h^{7} 1'=O_h^{7} \times \{1,1'\}$.
Since the orbital ordering is induced by IR of the zero wave vector
$k_{11}=0$,
the group-theoretical analysis can be performed by using representations of
the point groups $O_h$ and $O_h1'$, because they are isomorphic to the IR of
the Shubnikov group $O_h^{7}1'$.
The three-component OP $c$, describing the ferromagnetic ordering in the cubic
crystal of the class $O_h$, is transformed under the IR $T_{1g}$.
However, OP has negative parity with respect to the inversion of the magnetic
moment $1'$, which means that the true OP is the IR $T_{1g}$ of the point
magnetic group $O_h1'$
\cite{Izyumov1991}.
Thus, the expansion of the orbital magnetic moment of 2(a)-nodes has the
form
\begin{equation}
  \label{eq:m-c}
  \mathbf{m} = c_1 \mathbf{m}_1 + c_2 \mathbf{m}_2 + c_3 \mathbf{m}_3.
\end{equation}
Unlike the expansion (\ref{eq:m-A}), the above expansion is symmetrically
correct.
Transformations of a vector in space can be performed in two ways.
An ``active'' transformation means that the vector itself is rotated, while
the basis vectors remain fixed.
A ``passive'' transformation means that the vector is fixed and the basis
vectors are rotated around an axis, which goes trough the origin of the
coordinate frame.
In both cases the elements of the same group $O_h$ or $O_h1'$ are used.
The triplet $(A_7,A_5,A_2)$, which stems from the expansion of the reducible
representation $T_{2g} \otimes T_{2g}$ of the group $O_h$ \cite{Jirak_prb92},
transforms 
under the IR $T_{1g}$ of the group $O_h$, or, which is the same, under the IR
$T_{1g}$ of the group $O_h 1'$.
This follows from the properties of the density matrix, whose elements,
in the absence of the spin-orbital interaction, are insensitive to
the time inversion.
The OP $c$ and the triplet $(A_7,A_5,A_2)$ behave equally under the rotations
and the spatial inversion.
The properties of the OP $c$, which are caused by the proper rotations, can
also be attributed to the triplet $(A_7,A_5,A_2)$.
On the contrary, the properties of $c$, which are caused by the time
inversion, should not be attribute to the triplet $(A_7,A_5,A_2)$

\begin{figure}[htbp]
  \centering
  \includegraphics[width=0.6\textwidth]{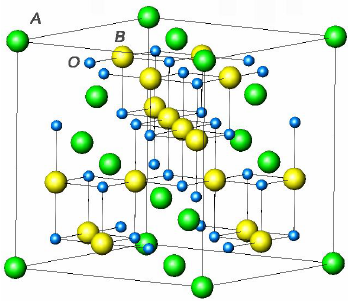}
  \caption{Structure of a normal spinel $\mathrm{A} \mathrm{B}_2
    \mathrm{O}_4$.
  Green spheres are 2(a)-position A-cations,
  yellow spheres are 4(c)-position B-cations,
  small blue spheres are 8(e)-position O-anions.}
  \label{fig:2}
\end{figure}


\section{ Magnetic symmetry of orbitally-ordered phases}
\label{sec:VI}

Transformational properties of the tensor parameter of a physical
phenomenon, which occurs below the temperature of the phase transition, should
comply with the symmetry of the emerging phase.
Namely, the tensor parameter should be invariant with respect to the
symmetry transformations induced by elements of the symmetry group of the
low-symmetric phase \cite{sakhnenko1986gm,stokes_isotropy}.
This leads to the equation
\begin{equation}
  \label{eq:U-c}
  \mathbf{\hat U}(g) \vc = \vc,
\end{equation}
where $g \in M_D$ with $M_D$ being the magnetic symmetry group of the
orbitally ordered phase, $\mathbf{\hat U}(g)$ is any of the generator matrices
of the critical IR.
The number of equations (\ref{eq:U-c}) is equal to the number of the
generators of the matrix group of IR.
If the vector $\vc$ satisfies equations for all generators, then it
describes the most symmetric phase.
The least symmetric phase corresponds to the case $\mathbf{\hat U}(g) =E$.
Geometrically this means that the vector $\vc$ is randomly oriented with
respect to the axes of the coordinate frame in the space of OP's.
Otherwise $\vc$ is oriented along some preferred directions in the space of
OP's.
In \cite{stokes_isotropy} the procedure of the determination of subgroups
using the equation 
(\ref{eq:U-c}) was applied to all 230 space groups.
Also, in \cite{stokes_isotropy} orientations of the critical OP,  satisfying
the Lifshits condition, 
as well their respective subgroups, were determined.
However, in our case we have to deal with Shubnikov groups and, in order to
use results of \cite{stokes_isotropy}, an additional analysis is necessary.
The corresponding details are given in Appendix~\ref{sec:gauge-invariance},
and the results are given in the table~\ref{tab:one}.

\begin{table}[ht]
  \centering
  \begin{tabular}{|c|c|c|c|c|c|c|}
\hline 
 $c$   & $ccc$ & $0cc$ & $c00$ & $c_1c_2c_2$ & $0c_1c_2$ & $c_1 c_2 c_3$ \\
\hline
 $M_D$   & $D_3^{7} (C_3^{4})$ & $D_2^{8}(C_2^{1})$ & $D_4^{5}(C_4^{5})$ &
                                                                           $C_2^{2}(C_1^{1})$
                                             & $C_2^{3}(C_1^{1})$ & $C_1^{1}$ \\
\hline
 $A_{2u}$   & $a$ & - & - & $a$ & - & $a$ \\
\hline
 $E_g$   & - & $0a$ & $a, - a \sqrt 3$ & $0,a$ & $a,b$ & $a,b$ \\
\hline
 $T_{2g}$  & $aaa$ & $0aa$ & & $abb$ & $0ab$ & $abc$ \\
    \hline
  \end{tabular}
  \caption{Low-symmetry orbitally ferromagnetic phases, induced by the critical OP
  $T_{1g}$ of the Shubnikov group $O_h^{7}1'$ and the related representations
  for the vector $k_{11}=0$.}
  \label{tab:one}
\end{table}

According to the table~\ref{tab:one}, there are six orbitally ordered phases,
which could possibly exist.
We postpone the interpretation of experimental results based on the
table~\ref{tab:one} till we clarify the role of the secondary OP's and their
accompanying effects.


\section{Related effects}
\label{sec:VII}

As is well-known [\cite{Izyumov1990,Borlakov_1999}, the equation of the kind
(\ref{eq:U-c}) is satisfied 
not only for the critical OP's, but also for some other additional OP's,
induced by other 
IR's, under the condition that the corresponding equations are symmetrically
compatible.
In this case, the additional OP's can have non-zero values in each of the six
possible phases.
Most important of the corresponding accompanying phenomena are the
deformations of the coordination polyhedron and the charge ordering on J-T
nodes.
The deformations are described by the symmetric deformation tensor $u_{ij}$.
From its diagonal components, $u_{ii}$, the two-dimensional OP can be built
\begin{equation}
  \label{eq:op-2}
  a_1 = \frac{1}{\sqrt 6} (2u _{zz} - u_{xx} - u_{yy}), \quad
  a_2 = \frac{1}{\sqrt 2} ( u_{yy} - u_{xx}),
\end{equation}
which describes the tensile-compressive deformations,
and the three-dimensional OP
\begin{equation}
  \label{eq:op-3}
  b_1 = u_{yz}, \quad   b_2 = u_{zx}, \quad  b_3 = u_{xy},
\end{equation}
which describes the shearing deformation.
Table~\ref{tab:2} provides with the information about the nodes of the spinel
lattice, which are displaced under the deformations, described by
 (\ref{eq:op-2}) and (\ref{eq:op-3}).

\begin{table}[ht]
  \centering
  \begin{tabular}{|l|c|c|c|}
    \hline
\hspace{2pc} Positions & 2(a) & 4(d) & 8(e) \\ 
 IR &  && \\ \hline
        11-4            & 0-1 & 1-0 & 1-1 \\ \hline
        11-5            & 0-0 & 0-0 & 1-0 \\ \hline
        11-7            & 1-0 & 1-0 & 2-1 \\ \hline
  \end{tabular}
  \caption{Multiplicity of the IR of the group $O_h^7$ for the wave vector
    $k_{11}=0$ in the mechanic and permutation representations [14].}
  \label{tab:2}
\end{table}

From the 3-rd row of the table~\ref{tab:2} it follows that under the
tensile-compressive deformation only the oxygen atoms are displaced (position
8(e)).
Shear deformations under the IR 11-7 lead to the displacement of all atoms of the
spinel structure.
From the 4-th row of the table~\ref{tab:one} and the 1-st row of the
table~\ref{tab:2} it follows that the charge ordering on positions 2(a) is
possible only in the trigonal, monoclinic or triclinic phases.

We will discuss the content of the tables~\ref{tab:one}-\ref{tab:2} in more
detail in the below section~\ref{sec:IX}, were the experimental data related
to spinels $\cu\chr_2\ox_4$ and $\nc\chr_2\ox_4$ are analyzed.


\section{Spin ordering}
\label{sec:VIII}

Spin ordering, which occurs at lower temperatures against the orbital
ordering, can be naturally described within the general group-theoretical scheme.
In order to do so, we assume, following \cite{Borlakov_1999}, that the spin
ordering is 
induced by the restriction of the critical IR of the paramagnetic phase
symmetry group onto its subgroup, describing the symmetry of the ordered phase.
In our case this is restriction $F^{(-)}_{1g}$ of the Shubnikov group
$O_h^71'$ onto its tetragonal subgroup $D_4^5(C_4^5)$.
An analysis shows that the restriction of the IR can be decomposed into a sum
$[T_{1g}^{(-)}]=A_1+E$, where $A_1$ is the unit IR of the group
$D_4^5(C_4^5)$, and $E$ is its two-dimensional IR.
Similarly, the restriction IR $T^{(+)}_{2g}$, describing the accompanying
magnetostriction, can be decomposed into the sum
$[T_{1g}^{(+)}]=B_2+E$, where $B_2$ is the one-dimensional IR of the
tetragonal group $D_4^5(C_4^5)$.
Above we have denoted the restrictions of IR's on the subgroup by enclosing
the notation of the IR of the starting group in square brackets.
The tetragonal group $C_{4v}^{11}$ is isomorphic to the Shubnikov group
$D_4^5(C_4^5)$, and its two-dimensional IR, $E$, of the wave
vector $k_{14}$, is isomorphic to the magnetic representation of the Shubnikov
group $D_4^5(C_4^5)$.

\begin{table}[ht]
  \centering
  \begin{tabular}{|l|c|c|c|}
    \hline
Critical OP    & (s,0) & (s,s) &  (s,s') \\  \hline
Deformation OP & (a,0) & (a,a) &  (a,b)  \\ \hline
Subgroups G    & 8 Cm  & 9 Cc  &  1 P1 \\ \hline
Subgroups M    & $C_2^3 (C_1^1)$ & $C_2^3 (C_1^1)$ & $C_1^1$ \\ \hline
  \end{tabular}
  \caption{ Possible spin-ordered states, induced by the critical OP
    of the Shubnikov group $D_4^5(C_4^5)$ $E$ of the wave vector
    $k_{14}=0$ \cite{sakhnenko1986gm}.}
  \label{tab:3}
\end{table}

The group-theoretical analysis, presented above, allows us to proceed to the
interpretation of the experimental data on the orbital, charge and spin
ordering in spinels $\cu\chr_2\ox_4$ and $\nc\chr_2\ox_4$.


\section{Analysis of experimental data}
\label{sec:IX}

The crystals under consideration are characterized by the following
temperatures of the orbital and spin ordering:
$\cu\chr_2\ox_4$ ($T_{orb}=310$ K, $T_c=70$ K) and $\nc\chr_2\ox_4$
($T_{orb}=860$ K, $T_c=135$ K)
\cite{Lenglet_1986,Kino_jpsj66,Suchomel_prb12}.
Orbitally ordered phase is tetragonally distorted and it is described by the
space group $I 4_1/amd=D_{4h}^{19}$ (No 141).
When the temperature decreases these crystals undergo the magnetic ordering;
interactions are somewhat distorted for $\nc \chr_2\ox_4$, but not for
$\cu\chr_2\ox_4$.
Further structural changes take place at the temperature of magnetic ordering.
They lead to the distortion of both materials into the rombic structure,
corresponding to the space group $D_{2h}^{24}=Fddd$ (No 70)
\cite{Lenglet_1986,Kino_jpsj66,Suchomel_prb12}.
$\ni\chr_2\ox_4$ exhibits additional distortions, possibly, in the same
space group, but at even lower transition temperature $T=30$ K
\cite{Suchomel_prb12}.
This is the essence of the experimental data on the structural changes in the
chromite spinels under consideration.
The unified theoretical model of the orbital, charge and spin ordering in
these compounds, which was presented above, predicts the following chain of
the magnetic symmetry transformations:
\begin{equation}
  \label{eq:m-sym-1}
  O_h^7 1' \xrightarrow[310 \, \mathrm{K}]{}
  D_4^5 (C_4^5)  \xrightarrow[70 \, \mathrm{K}]{}
  C_2^3 (C_1^1) \xrightarrow[30 \, \mathrm{K}]{} C_1^1.
\end{equation}
If we remove dashes from the generators of the magnetic groups, then we will
obtain the chain of space groups
\begin{equation}
  \label{eq:m-sym-2}
  O_h^7 \xrightarrow[310 \, \mathrm{K}]{}
  D_4^5 \xrightarrow[70 \, \mathrm{K}]{}
  C_2^3 \xrightarrow[30 \, \mathrm{K}]{} C_1^1.
\end{equation}
However, this chain predicts much less symmetric phases, than those experimentally
discovered. 
Nevertheless, there is an agreement between the chains
(\ref{eq:m-sym-1}),~(\ref{eq:m-sym-2}) and the chain
\begin{equation}
  \label{eq:m-sym-exp}
  O_h^7 \xrightarrow[310 \, \mathrm{K}]{}
  D_{4h}^{19} \xrightarrow[70 \, \mathrm{K}]{}
  D_{2h}^{24} \xrightarrow[30 \, \mathrm{K}]{} G_x,
\end{equation}
which corresponds to the experimental data.
This agreement stems from the fact that the observed crystal structure is
affected by the influence of the secondary OP's (\ref{eq:op-2}),
(\ref{eq:op-3}), built from the components of the deformation tensor.
In close vicinity of the phase transition
the displacements, caused by these OP's, are small comparing to the critical
OP.
However, they increase as the temperature moves away from the phase
transition.
Accordingly, the secondary OP's affect the crystal structure more than the
critical OP.
Consequently, orientations of deformation OP's in the space of ordering
parameters, which are symmetrically matched with the orientation of the
critical OP, correspond to the experimentally observed chain of the
structural changes (\ref{eq:m-sym-exp}).
Simultaneously, the change of the true magnetic symmetry corresponds to the
transformation chain (\ref{eq:m-sym-1}).
This can be seen from the general theory of deformation transitions, presented
in \cite{Sakhnenko1980,Sakhnenko1979}.
The possible transformations for the tensile deformations are exactly those,
given by the chain (\ref{eq:m-sym-exp}).
For the shearing deformation the low-symmetric phases
$O_h^7 \to D_{3d}^5, D_{2h}^{28}, C_{2h}^3, C_1^1$ are possible.
Deformations caused by the shearing occur below the point of spin ordering
and the phase is not $D_{2h}^{24}$, but $D_{2h}^{28}$, which matches with
the shearing deformations of the kind $a=(a,a)$ corresponding to the spin
moment $s=(s,s)$ emerging in the basis plane of the tetragonal phase.
Unfortunately, there is no information on the charge ordering in the crystals
we consider.
There is no charge ordering in the tetragonal phase due to the symmetry
arguments, and it only emerges in one of the two possible monoclinic phases.
However, the monoclinic phase arises already at the spin ordering and it is
seen from the distortions of the anion base as a rombic phase.
In this ``rombic phase'' one could expect the splitting of the integrality of
the electron charge on 2(a) nodes of the tetragonal sublattice.


\section{Conclusion}
\label{sec:concl}

Results of the present work can be summarized as follows.
\begin{enumerate}
\item 
We have shown, for the first time, that orbital degrees of freedom, introduced
into the Landau theory of phase transitions in the work \cite{Gurin_jetp89},
are the source of orbital magnetism.

\item 
The symmetry of an orbitally disordered state is accurately
described by the gray Shubnikov paramagnetic group, which is $O_h^7 1'$ in the
case of spinels.
The critical ordering parameter, responsible for the chain of phase
transitions and the corresponding chain of symmetry groups, realized on the
thermodynamic path of the crystal, is a magnetic one.

\item 
The removal of the orbital degeneracy is caused by the Coulomb interaction, or,
more precisely, its exchange part through superexchange.
It clearly follows from the fact that elements of the Bethe crystal field
theory were incorporated into our calculation scheme.
If, instead of Bethe theory, the ligand field theory will be used, then the
exchange integrals will explicitly appear in the calculations.
This will drastically complicate the calculations as the density matrix in
this case
contains coordinates of at least three ions: J-T-cation--anion--J-T-cation.
Correspondingly, the wave functions will depend on three radius-vectors.
As for the displacements of anions of the coordination polyhedron, then this is
simply a consequence of the cooperative Jahn-Teller effect 
(orbital ordering in current terms),
caused by superexchange.

\item 
Coefficients in the equation (\ref{eq:m-A}), connecting mean orbital current
with the basis currents, do not form the critical ordering parameter
since they transform differently under the time inversion.
Nevertheless, both triplets behave equally under the proper rotations.
We have ``decoupled'' the ordering parameter from its origin, i.e.,
elements of the density matrix.
This gives the possibility to describe arbitrary phase transitions causing the
orbital ordering, i.e., not only those happening without the change of the
translation symmetry.

\item 
The spin ordering is naturally included into the general group-theoretical
scheme of the subordination of phases through the subordination of their symmetry
groups to the symmetry group of the parent phase.
This scheme is a generalization of Landau theory and it was already applied by
one of authors to the theory of phase transitions in $\cu\fe_2\ox_4$
\cite{Borlakov_1999}.
For copper ferrite $CuFe_2O_4$, $T_{\text{orb}}=631\, \mathrm{K}$, $T_c=730\, \mathrm{K}$, and spin ordering precedes orbital ordering. Both transitions are described by a unified theoretical group model with a scheme of subordination of symmetry groups of the newly emerging phases.
In this scheme the new ordering parameter is caused by the restriction of the
IR of the symmetry group of the parent phase upon its subgroup, from which the
phase transition is started.

\item 
The symmetry of theoretically established magnetic groups of low-symmetric
phases, as a rule, is below the symmetry of crystal lattices of those phases.
The reason is that the symmetry of crystal lattices is defined by
displacements of atoms (mostly by anions), and it is higher than the real
symmetry described by the magnetic ordering parameter.
The determination of the real magnetic symmetry requires more elaborate ways
than the conventional X-ray crystallography.

\item 
To ascertain the magnetic nature of the phase transition in the point of
orbital ordering it would be sufficient to measure the temperature dependence
of the magnetic susceptibility.

\item 
Although the orbital ordering is determined by the superexchange interaction,
the magnetic phases are anisotropic.
This is due to the fact that atomic orbitals are built into the lattice
and Coulomb forces do not allow for the free orientation of orbital currents
with respect to the crystallographic axes.
If the orbital ordering takes place at a lower temperature than the spin
ordering, as for $\cu\fe_2\ox_4$, then it is possible to observe
experimentally the spin ordered isotropic ferromagnetic phase
\cite{Borlakov_1999}.

\item
Spin ordering in chromite spinels, considered in this work, is caused by the
spin-orbital interaction.
This is confirmed by rather low temperatures of the transition into the
spin ordered phase.
\end{enumerate}

In literature much attention is paid towards the question on the mechanism of
the orbital ordering.
From the above consideration it follows that there are two mechanisms.
Under the transition from the disordered phase to the orbitally ordered phase
the mechanism is the superexchange by Kugel-Khomsky \cite{Kugel_ufn82}.
When the orbital ordered happens against the background of the spin ordering,
then the mechanism is the spin-orbital interaction, which gives rise to
the orbital contribution into the spin magnetism and causes the 
emergence of the magnetic anisotropy.
The starting idea of Jahn and Teller for the cooperative effect is not
correct since the displacements of anions do not lead to the removal of
degeneracy, which is just the result of orbital ordering.
The manifestation of displacements when moving away from the point of
orbital ordering leads to an illusion that displacements define
the symmetry of the crystal lattice.
This is a misconception, though it is supported by the X-ray crystallography.
The determination of the true magnetic symmetry requires more elaborate
experimental techniques.

On the visualization tools used in the experimental and theoretical
studies of orbital ordering.
Here, the prevailing method of the interpretation of the initial and orbitally
ordered states of a J-T node is based on the images of 3d-orbitals.
However, these orbitals correspond to purely quantum mechanical states of an
ion subjected to thermal action of the environment, which cannot be correctly
taken into account within the framework of model theories.
The correct account for the environmental influence on a J-T node can be given
by the nodal density matrix, which we used to compose the basis of
averaged currents.
Images of averaged currents are the accurate way to analyze the processes on a
J-T node and its surroundings.

Finally, we note that the search for the manifestation of orbital magnetism
and its experimental measurements are carried out rather intensively \cite{CECAM}.
Nevertheless, we have not seen attempts to find the orbital magnetism in J-T
crystals  so far.

\textbf{Acknowledgement.}
The authors would like to thank Alexey Meremyanin for helpful advice and assistance in preparing the manuscript.

\appendix 


\section{Calculations of the basis magnetic moments and the
proof of the gauge invariance of the basis currents}
\label{sec:m-moments-gauge-inv}


Starting from the explicit expressions (\ref{eq:phi-s}) for the basis
functions $\varphi_i$, we can calculate the basis current
$\mathbf{J}_3$ according to  (\ref{eq:J-s}), (\ref{eq:def-J}):
\begin{equation}
  \label{eq:J3}
  \mathbf{J}_3 = \kappa^{2} \frac{e\hbar}{m} \cdot
  \frac{y z^{2}\, \mathbf{i} - x z^{2} \, \mathbf{j}}{r^{4}}.
\end{equation}
Let us calculate the magnetic moment $\mathbf{m}_3$, induced by the basis
current $\mathbf{J}_3$.
We begin with the expression \cite{Landau_T8,Hirst_rmp97}:
\begin{equation}
  \label{eq:m-int-a}
    \vm = \int \mathbf{M}(\vvr) \, d \vvr 
= \int \vvr \times \mathbf{J}(\vvr) \, d \vvr.
\end{equation}
Everywhere above we have ignored the presence of the square of the radial wave
function $R_{32}(r)$ in the orbital current.
However, upon the substitution of $\vcj_3$ from (\ref{eq:def-J}) into
(\ref{eq:m-int-a}) the current has to be multiplied with $R^{2}_{32}(r)$.
As a result, we obtain,
\begin{equation}
  \label{eq:ms-es}
  \vm_1 = \frac{e \hbar}{m} \tilde{\ve}_1, \quad
  \vm_2 = \frac{e \hbar}{m} \tilde{\ve}_2, \quad
  \vm_3 = \frac{e \hbar}{m} \tilde{\ve}_3,
\end{equation}
where $\tilde{\ve}_i$ are the axial Cartesian basis vectors whose parity with
respect to the time inversion is equal to $(-1)$
\cite{Sirotin_Shask_92}.
Thus, the three magnetic moments are directed along the axes of the coordinate
frame and have the same magnitude, which is twice the Bohr magneton.


The stationary current $\vcj$ is connected with the magnetization vector
$\vcm$ by \cite{Balcar_1975}:
\begin{equation}
  \label{eq:J-M}
  \vcj(\vvr) = \mathrm{rot}\, \vcm(\vvr).
\end{equation}
Integrating this equation we obtain magnetization caused by the stationary
current:
\begin{equation}
  \label{eq:M-J}
  \vcm(\vvr) = \vvr \times \vcj(\vvr) + \nabla g(\vvr),
\end{equation}
where $g(\vvr)$ is an arbitrary function, which is determined by the gauge
convention.
Usually, the Trammel gauge is used \cite{Balcar_1975,Hirst_rmp97}.
Following this convention, the magnetization vector becomes
\begin{equation}
  \label{eq:Trm-gauge}
  \vcm(\vvr) = \int_1^{\infty} \lambda \vvr \times \vcj(\lambda \vvr) \,
  d \lambda.  
\end{equation}
As is shown in \cite{Balcar_1975,Hirst_rmp97}, equation (\ref{eq:Trm-gauge})
allows one to express the orbital magnetization in the state of thermal
equilibrium via the average of the orbital momentum density operator:
\begin{equation}
  \label{eq:M-l}
  \vcm(\vvr) = \frac{e \hbar}{m}
\int_1^{\infty} \lambda \vvr \times \vl(\lambda \vvr) \, d \lambda.    
\end{equation}
In fact, this equation is a justification of the often used in literature
assumption that the ion of the crystal lattice in a triply degenerated
orbital state has an effective orbital momentum quantum number $l=1$.
Let us analyze whether the basis currents satisfy the Trammel gauge.
Substituting the explicit expression for $\vcj_3$ from (\ref{eq:def-J})  into
(\ref{eq:Trm-gauge}) and noting the expression for the radial wave function
$R_{32}(r) = N (r/a_0)^{2} \exp ( - r/(3a_0))$, where $a_0$ is the Bohr
radius, we obtain
\begin{equation}
  \label{eq:M3}
  \vcm_3(\vvr) = \int_0^{\infty} \lambda \vvr \times \vcj_3(\lambda \vvr)\,
  R^{2}_{32} (\lambda r) \, d \lambda
  = \vcm_3(\vvr) e^{\frac{2r}{3a_0}}
  \left( \frac{3a_0}{2r} \right)^{3}
  \int_{2r/(3a_0)}^{\infty} x^{2} e^{-x} d x,
\end{equation}
where the variable change $2\lambda r/(3a_0)=x$ has been applied.
Calculating the integral on the rhs of (\ref{eq:M3}), we obtain the factor at
$\vcm_3$, which should be equal to unity:
\begin{equation}
  \label{eq:x}
  \frac{1}{x^{3}} (x^{2} + 2x +2 ) =1, \quad x= \frac{2r}{3 a_0}.
\end{equation}
This equation is equivalent to the cubic equation for $x$:
\begin{equation}
  \label{eq:x-cube}
  x^{3} -x^{2} - 2x -2 =0.
\end{equation}
This equation has one real root $x_1 \approx 2.269530842$.
The use of the Trammel gauge localizes $\vcm(\vvr)$ for large $r$
 \cite{Balcar_1975,Hirst_rmp97}.
The cutoff radius is connected with the root $x_1$ of (\ref{eq:x-cube}) by
\begin{equation}
  \label{eq:r-x}
  r_1 = x_1 \frac{3a_0}{2} \approx 2.269530842 \frac{3a_0}{2}
  \approx 3.4\, a_0.
\end{equation}


\section{On the connection of  irreducible representations of
  Shubnikov and Fedorov groups}
\label{sec:gauge-invariance}

Mutual relations of representations of Shubnikov groups are considered in the
book \cite{Izyumov1991}, which we will follow in the below consideration.
We start from the so-called ``gray'' Shubnikov groups.
They can always be presented in the form
\begin{equation}
  \label{eq:shu-1}
  M = G 1' = G + 1' \times G.
\end{equation}
Groups of the kind (\ref{eq:shu-1}) do not allow for the magnetic ordering and
they describe the symmetry of paramagnetic phases.
From their definition it becomes evident that the number of IR's for a given
Shubnikov group is twice than that for the corresponding Fedorov group.
Instead of one IR $\tau_\mu$ of the group $G$, we obtain two IR's of the group
$G 1'$, one of which, ${\tau^e}_\mu$, is even with respect to the time
inversion, and the second one, ${\tau^o}_\mu$, is odd.
In the representation $\tau_\mu^e$ the element $1'$ is associated with the
unit matrix, and in the representation $\tau_\mu^o$ $1'$ is associated
with the unit matrix with the minus sign.

All Fedorov groups can be divided into two groups -- centrosymmetric and
non-centrosymmetric. 
Centrosymmetric groups can be represented in a form similar to
(\ref{eq:shu-1}):
\begin{equation}
  \label{eq:shu-2}
  G = G_0 + I \times G_0,
\end{equation}
where $G_0$ is the non-centrosymmetric subgroup of the group $G$, and $I$ is
the space inversion.

If $\tau_\mu$ is considered to be the IR of $G_0$, then IR's of $G$ are
obtained by doubling with respect to the parity property $I$ in the same
way as was described above for the magnetic group. 
Instead of $\tau_\mu$ we have two representations: $\tau_{\mu+}$ and
$\tau_{\mu -}$, which are even and odd with respect to the space
inversion, respectively.
In these notations the OP $c$ is transformed under the IR $\tau_{\mu+}$.
If we compare the matrices of the IR's $\tau_{\mu+}$ and $\tau_{\mu-}$, then
we will see that these IR's are isomorphic as matrix groups, while being
different in a physical sense.
The polar vector $\vvr$ is transformed under the IR $\mu_{\mu-}$.
In the case of phase transitions under the IR of the vector $k=0$, which takes
place in the case of orbital ordering in the spinels under study, then this is
the IR $T_{1 u}$.
The symmetry of ordered phases, induced by the IR $T_{1u}$, is well known and
is presented in a number of sources \cite{sakhnenko1986gm}.
In the table~\ref{tab:b1} the possible orientations of
the OP in the image space and their corresponding Fedorov groups describing
the symmetry of ordered phases are given.

\begin{table}[htbp]
  \centering
  \begin{tabular}{|c|c|c|c|c|c|c|}
\hline
   c & ccc & 0cc & c00 & $c_1c_2c_2$ & $0c_1c_2$ & $c_1c_2c_3$ \\
\hline
 $G_D$   & $C_{3v}^{5}$ & $C_{2v}^{20}$ & $C_{4v}^{11}$ & $C_{s}^{3}$ &
                                         $C_{s}^{4}$  & $C_{1}^{1}$  \\
\hline
  $M_D$  & $D_3^{7} (C_{3}^{4})$ & $D_2^{8} (C_{2}^{1})$ &
   $D_4^{5} (C_{4}^{5})$ & $C_2^{2} (C_{1}^{1})$ & $C_2^{3} (C_{1}^{1})$
                      & $C_1^{1}$ \\
\hline
  \end{tabular}
  \caption{ }
  \label{tab:b1}
\end{table}

$G_D$ groups are formal solutions, which can be used to determine the true
magnetic symmetry groups.
As an example let us consider the tetragonal group $G_D=C_{4v}^{11}$.
Generators of this group are given in table~\ref{tab:b2} according to Kovalev
\cite{Kovalev_1993}. 

\begin{table}[htbp]
  \centering
  \begin{tabular}{|c|c|c|c|c|}
\hline
  $g_i$ & 14,15 & 4 & 26,27 & 37,40 \\
\hline
 $C_{4v}^{11}$   & $01\frac{1}{2}$ & 000 & 000 & $01\frac{1}{2}$  \\
\hline
 $g_i$  & 14,15 & 4 & $2',3'$ & $13',16'$ \\
\hline
 $D_4^{5}$  & 001 & 000 & 000 & 001 \\
\hline
 $g_i$  & 14,15 & 4 &  &  \\
\hline
 $C_4^{5}$  & 001 & 000 &  &  \\
\hline
  \end{tabular}
  \caption{ }
  \label{tab:b2}
\end{table}

Next, we transform the rotoinversion part of the element
$g_i$ according to the rule $h_{i +24} =I h_i$ and assign it the dash
$I h_i → h'_i$, i.e. replace the spatial inversion with the inversion of the
magnetic moment and declare it as a magnetic group generator.
With the dashes removed we have space group with generators
$(h_2|000)$, $(h_3|000)$, $(h_4|000)$ and
$(h_{13}|001)$, $(h_{14}|001)$, $(h_{15}|001)$, $(h_{15}|001)$.
The mismatch of accompanying translations should not lead to confusion, since
these translations cannot be determined from the connection of the orientation
of the OP with the corresponding subgroups of the initial group.
Nevertheless, the lack of accompanying translations should be preserved, as
well as the relation between them when these exist.

Thus, in Sch\"onflies notations the magnetic symmetry group is denoted as
$D_4^{5}(C_4^{5})$; its generators are elements given in the 3-rd and 4-th
rows of the table~\ref{tab:b2}.
Group $C_4^{5}$ defined by unprimed generators is the index 2
subgroup for the group $D_4^{5}$.
Exacty in this way the Shubnikov magnetic groups are denoted \cite{Landau_T8}
as the two-position symbol composed of two Fedorov groups.
Although our derivation of magnetic groups slightly differs
from that recommended in the book \cite{Izyumov1991}, the result is in full
agreement with the Indenbom-Niggli theorem.


\begin{thebibliography}{23}%
\makeatletter
\providecommand \@ifxundefined [1]{%
 \@ifx{#1\undefined}
}%
\providecommand \@ifnum [1]{%
 \ifnum #1\expandafter \@firstoftwo
 \else \expandafter \@secondoftwo
 \fi
}%
\providecommand \@ifx [1]{%
 \ifx #1\expandafter \@firstoftwo
 \else \expandafter \@secondoftwo
 \fi
}%
\providecommand \natexlab [1]{#1}%
\providecommand \enquote  [1]{``#1''}%
\providecommand \bibnamefont  [1]{#1}%
\providecommand \bibfnamefont [1]{#1}%
\providecommand \citenamefont [1]{#1}%
\providecommand \href@noop [0]{\@secondoftwo}%
\providecommand \href [0]{\begingroup \@sanitize@url \@href}%
\providecommand \@href[1]{\@@startlink{#1}\@@href}%
\providecommand \@@href[1]{\endgroup#1\@@endlink}%
\providecommand \@sanitize@url [0]{\catcode `\\12\catcode `\$12\catcode
  `\&12\catcode `\#12\catcode `\^12\catcode `\_12\catcode `\%12\relax}%
\providecommand \@@startlink[1]{}%
\providecommand \@@endlink[0]{}%
\providecommand \url  [0]{\begingroup\@sanitize@url \@url }%
\providecommand \@url [1]{\endgroup\@href {#1}{\urlprefix }}%
\providecommand \urlprefix  [0]{URL }%
\providecommand \Eprint [0]{\href }%
\providecommand \doibase [0]{https://doi.org/}%
\providecommand \selectlanguage [0]{\@gobble}%
\providecommand \bibinfo  [0]{\@secondoftwo}%
\providecommand \bibfield  [0]{\@secondoftwo}%
\providecommand \translation [1]{[#1]}%
\providecommand \BibitemOpen [0]{}%
\providecommand \bibitemStop [0]{}%
\providecommand \bibitemNoStop [0]{.\EOS\space}%
\providecommand \EOS [0]{\spacefactor3000\relax}%
\providecommand \BibitemShut  [1]{\csname bibitem#1\endcsname}%
\let\auto@bib@innerbib\@empty
\bibitem [{\citenamefont {Gurin}\ \emph {et~al.}(1989)\citenamefont {Gurin},
  \citenamefont {Budrina},\ and\ \citenamefont {Syromyatnikov}}]{Gurin_jetp89}%
  \BibitemOpen
  \bibfield  {author} {\bibinfo {author} {\bibfnamefont {O.~V.}\ \bibnamefont
  {Gurin}}, \bibinfo {author} {\bibfnamefont {G.~L.}\ \bibnamefont {Budrina}},\
  and\ \bibinfo {author} {\bibfnamefont {V.~N.}\ \bibnamefont
  {Syromyatnikov}},\ }\href@noop {} {\bibfield  {journal} {\bibinfo  {journal}
  {JETP}\ }\textbf {\bibinfo {volume} {68}},\ \bibinfo {pages} {770} (\bibinfo
  {year} {1989})}\BibitemShut {NoStop}%
\bibitem [{\citenamefont {Izyumov}\ and\ \citenamefont
  {Syromyatnikov}(1998)}]{Izyumov_pt98}%
  \BibitemOpen
  \bibfield  {author} {\bibinfo {author} {\bibfnamefont {Y.~A.}\ \bibnamefont
  {Izyumov}}\ and\ \bibinfo {author} {\bibfnamefont {V.~N.}\ \bibnamefont
  {Syromyatnikov}},\ }\href {https://doi.org/10.1080/01411599808222118}
  {\bibfield  {journal} {\bibinfo  {journal} {Phase Transitions}\ }\textbf
  {\bibinfo {volume} {66}},\ \bibinfo {pages} {23} (\bibinfo {year}
  {1998})}\BibitemShut {NoStop}%
\bibitem [{\citenamefont {Jir\'ak}(1992)}]{Jirak_prb92}%
  \BibitemOpen
  \bibfield  {author} {\bibinfo {author} {\bibfnamefont {Z.}~\bibnamefont
  {Jir\'ak}},\ }\href {https://doi.org/10.1103/PhysRevB.46.8725} {\bibfield
  {journal} {\bibinfo  {journal} {Phys. Rev. B}\ }\textbf {\bibinfo {volume}
  {46}},\ \bibinfo {pages} {8725} (\bibinfo {year} {1992})}\BibitemShut
  {NoStop}%
\bibitem [{\citenamefont {Landau}\ and\ \citenamefont
  {Lifshitz}(1984)}]{Landau_T8}%
  \BibitemOpen
  \bibfield  {author} {\bibinfo {author} {\bibfnamefont {L.~D.}\ \bibnamefont
  {Landau}}\ and\ \bibinfo {author} {\bibfnamefont {E.~M.}\ \bibnamefont
  {Lifshitz}},\ }\href@noop {} {\emph {\bibinfo {title} {Electrodynamics of
  Continuous Media}}},\ \bibinfo {edition} {2nd}\ ed.,\ Vol.~\bibinfo {volume}
  {8}\ (\bibinfo  {publisher} {Pergamon},\ \bibinfo {year} {1984})\BibitemShut
  {NoStop}%
\bibitem [{\citenamefont {Streltsov}\ and\ \citenamefont
  {Khomskii}(2017)}]{Streltsov_2017}%
  \BibitemOpen
  \bibfield  {author} {\bibinfo {author} {\bibfnamefont {S.~V.}\ \bibnamefont
  {Streltsov}}\ and\ \bibinfo {author} {\bibfnamefont {D.~I.}\ \bibnamefont
  {Khomskii}},\ }\href {https://doi.org/10.3367/UFNe.2017.08.038196} {\bibfield
   {journal} {\bibinfo  {journal} {Phys. Usp.}\ }\textbf {\bibinfo {volume}
  {60}},\ \bibinfo {pages} {1121} (\bibinfo {year} {2017})}\BibitemShut
  {NoStop}%
\bibitem [{\citenamefont {Rumer}\ and\ \citenamefont
  {Fet}(1970)}]{rumer_unit_sym}%
  \BibitemOpen
  \bibfield  {author} {\bibinfo {author} {\bibfnamefont {Y.~B.}\ \bibnamefont
  {Rumer}}\ and\ \bibinfo {author} {\bibfnamefont {A.~I.}\ \bibnamefont
  {Fet}},\ }\href@noop {} {\emph {\bibinfo {title} {Theory of unitary
  symmetry}}}\ (\bibinfo  {publisher} {Nauka},\ \bibinfo {address} {Moscow},\
  \bibinfo {year} {1970})\ \bibinfo {note} {(in Russian)}\BibitemShut {NoStop}%
\bibitem [{\citenamefont {Balcar}(1975)}]{Balcar_1975}%
  \BibitemOpen
  \bibfield  {author} {\bibinfo {author} {\bibfnamefont {E.}~\bibnamefont
  {Balcar}},\ }\href {https://doi.org/10.1088/0022-3719/8/10/014} {\bibfield
  {journal} {\bibinfo  {journal} {Journal of Physics C: Solid State Physics}\
  }\textbf {\bibinfo {volume} {8}},\ \bibinfo {pages} {1581} (\bibinfo {year}
  {1975})}\BibitemShut {NoStop}%
\bibitem [{\citenamefont {Landau}\ and\ \citenamefont
  {Lifshitz}(1991)}]{Landau_T3}%
  \BibitemOpen
  \bibfield  {author} {\bibinfo {author} {\bibfnamefont {L.~D.}\ \bibnamefont
  {Landau}}\ and\ \bibinfo {author} {\bibfnamefont {E.~M.}\ \bibnamefont
  {Lifshitz}},\ }\href@noop {} {\emph {\bibinfo {title} {Quantum Mechanics}}},\
  \bibinfo {edition} {3rd}\ ed.,\ Vol.~\bibinfo {volume} {3}\ (\bibinfo
  {publisher} {Pergamon},\ \bibinfo {year} {1991})\BibitemShut {NoStop}%
\bibitem [{\citenamefont {Kovalev}(1993)}]{Kovalev_1993}%
  \BibitemOpen
  \bibfield  {author} {\bibinfo {author} {\bibfnamefont {O.~V.}\ \bibnamefont
  {Kovalev}},\ }\href@noop {} {\emph {\bibinfo {title} {Representation of
  Crystallographic Space Groups}}}\ (\bibinfo  {publisher} {Taylor \&
  Francis},\ \bibinfo {year} {1993})\BibitemShut {NoStop}%
\bibitem [{\citenamefont {Lenglet}\ \emph {et~al.}(1986)\citenamefont
  {Lenglet}, \citenamefont {{d'Huysser}}, \citenamefont {Arsene}, \citenamefont
  {Bonnelle},\ and\ \citenamefont {Jorgensen}}]{Lenglet_1986}%
  \BibitemOpen
  \bibfield  {author} {\bibinfo {author} {\bibfnamefont {M.}~\bibnamefont
  {Lenglet}}, \bibinfo {author} {\bibfnamefont {A.}~\bibnamefont
  {{d'Huysser}}}, \bibinfo {author} {\bibfnamefont {J.}~\bibnamefont {Arsene}},
  \bibinfo {author} {\bibfnamefont {J.~P.}\ \bibnamefont {Bonnelle}},\ and\
  \bibinfo {author} {\bibfnamefont {C.~K.}\ \bibnamefont {Jorgensen}},\ }\href
  {https://doi.org/10.1088/0022-3719/19/17/003} {\bibfield  {journal} {\bibinfo
   {journal} {Journal of Physics C: Solid State Physics}\ }\textbf {\bibinfo
  {volume} {19}},\ \bibinfo {pages} {L363} (\bibinfo {year}
  {1986})}\BibitemShut {NoStop}%
\bibitem [{\citenamefont {Kino}\ and\ \citenamefont
  {Miyahara}(1966)}]{Kino_jpsj66}%
  \BibitemOpen
  \bibfield  {author} {\bibinfo {author} {\bibfnamefont {Y.}~\bibnamefont
  {Kino}}\ and\ \bibinfo {author} {\bibfnamefont {S.}~\bibnamefont
  {Miyahara}},\ }\href {https://doi.org/10.1143/JPSJ.21.2732} {\bibfield
  {journal} {\bibinfo  {journal} {Journal of the Physical Society of Japan}\
  }\textbf {\bibinfo {volume} {21}},\ \bibinfo {pages} {2732} (\bibinfo {year}
  {1966})}\BibitemShut {NoStop}%
\bibitem [{\citenamefont {Suchomel}\ \emph {et~al.}(2012)\citenamefont
  {Suchomel}, \citenamefont {Shoemaker}, \citenamefont {Ribaud}, \citenamefont
  {Kemei},\ and\ \citenamefont {Seshadri}}]{Suchomel_prb12}%
  \BibitemOpen
  \bibfield  {author} {\bibinfo {author} {\bibfnamefont {M.~R.}\ \bibnamefont
  {Suchomel}}, \bibinfo {author} {\bibfnamefont {D.~P.}\ \bibnamefont
  {Shoemaker}}, \bibinfo {author} {\bibfnamefont {L.}~\bibnamefont {Ribaud}},
  \bibinfo {author} {\bibfnamefont {M.~C.}\ \bibnamefont {Kemei}},\ and\
  \bibinfo {author} {\bibfnamefont {R.}~\bibnamefont {Seshadri}},\ }\href
  {https://doi.org/10.1103/PhysRevB.86.054406} {\bibfield  {journal} {\bibinfo
  {journal} {Phys. Rev. B}\ }\textbf {\bibinfo {volume} {86}},\ \bibinfo
  {pages} {054406} (\bibinfo {year} {2012})}\BibitemShut {NoStop}%
\bibitem [{\citenamefont {Izyumov}\ \emph {et~al.}(1991)\citenamefont
  {Izyumov}, \citenamefont {Naish},\ and\ \citenamefont
  {Ozerov}}]{Izyumov1991}%
  \BibitemOpen
  \bibfield  {author} {\bibinfo {author} {\bibfnamefont {Y.~A.}\ \bibnamefont
  {Izyumov}}, \bibinfo {author} {\bibfnamefont {V.~E.}\ \bibnamefont {Naish}},\
  and\ \bibinfo {author} {\bibfnamefont {R.~P.}\ \bibnamefont {Ozerov}},\
  }\href {https://doi.org/10.1007/978-1-4615-3658-1} {\emph {\bibinfo {title}
  {Neutron Diffraction of Magnetic Materials}}}\ (\bibinfo  {publisher}
  {Springer {US}},\ \bibinfo {year} {1991})\BibitemShut {NoStop}%
\bibitem [{\citenamefont {Sakhnenko}\ \emph {et~al.}(1986)\citenamefont
  {Sakhnenko}, \citenamefont {Talanov},\ and\ \citenamefont
  {Chechin}}]{sakhnenko1986gm}%
  \BibitemOpen
  \bibfield  {author} {\bibinfo {author} {\bibfnamefont {V.~P.}\ \bibnamefont
  {Sakhnenko}}, \bibinfo {author} {\bibfnamefont {V.~M.}\ \bibnamefont
  {Talanov}},\ and\ \bibinfo {author} {\bibfnamefont {G.~M.}\ \bibnamefont
  {Chechin}},\ }\href@noop {} {\bibfield  {journal} {\bibinfo  {journal} {Fiz.
  Met. Metalloved}\ }\textbf {\bibinfo {volume} {62}},\ \bibinfo {pages} {847}
  (\bibinfo {year} {1986})}\BibitemShut {NoStop}%
\bibitem [{\citenamefont {Stokes}\ and\ \citenamefont
  {Hatch}()}]{stokes_isotropy}%
  \BibitemOpen
  \bibfield  {author} {\bibinfo {author} {\bibfnamefont {H.~T.}\ \bibnamefont
  {Stokes}}\ and\ \bibinfo {author} {\bibfnamefont {D.~M.}\ \bibnamefont
  {Hatch}},\ }\href {http://stokes.byu.edu/iso/isotropy.html} {\emph {\bibinfo
  {title} {ISOTROPY}}}\BibitemShut {NoStop}%
\bibitem [{\citenamefont {Izyumov}\ and\ \citenamefont
  {Syromyatnikov}(1990)}]{Izyumov1990}%
  \BibitemOpen
  \bibfield  {author} {\bibinfo {author} {\bibfnamefont {Y.~A.}\ \bibnamefont
  {Izyumov}}\ and\ \bibinfo {author} {\bibfnamefont {V.~N.}\ \bibnamefont
  {Syromyatnikov}},\ }\href {https://doi.org/10.1007/978-94-009-1920-4} {\emph
  {\bibinfo {title} {Phase Transitions and Crystal Symmetry}}}\ (\bibinfo
  {publisher} {Springer Netherlands},\ \bibinfo {year} {1990})\BibitemShut
  {NoStop}%
\bibitem [{\citenamefont {Borlakov}(1999)}]{Borlakov_1999}%
  \BibitemOpen
  \bibfield  {author} {\bibinfo {author} {\bibfnamefont {K.~S.}\ \bibnamefont
  {Borlakov}},\ }\href@noop {} {\bibfield  {journal} {\bibinfo  {journal}
  {Physics of Metals and Metallography}\ } (\bibinfo {year}
  {1999})}\BibitemShut {NoStop}%
\bibitem [{\citenamefont {Sakhnenko}\ and\ \citenamefont
  {Talanov}(1980)}]{Sakhnenko1980}%
  \BibitemOpen
  \bibfield  {author} {\bibinfo {author} {\bibfnamefont {V.~P.}\ \bibnamefont
  {Sakhnenko}}\ and\ \bibinfo {author} {\bibfnamefont {V.~M.}\ \bibnamefont
  {Talanov}},\ }\href@noop {} {\bibfield  {journal} {\bibinfo  {journal} {Sov.
  Phys. Solid State}\ }\textbf {\bibinfo {volume} {22}},\ \bibinfo {pages}
  {458} (\bibinfo {year} {1980})}\BibitemShut {NoStop}%
\bibitem [{\citenamefont {Sakhnenko}\ and\ \citenamefont
  {Talanov}(1979)}]{Sakhnenko1979}%
  \BibitemOpen
  \bibfield  {author} {\bibinfo {author} {\bibfnamefont {V.~P.}\ \bibnamefont
  {Sakhnenko}}\ and\ \bibinfo {author} {\bibfnamefont {V.~M.}\ \bibnamefont
  {Talanov}},\ }\href@noop {} {\bibfield  {journal} {\bibinfo  {journal} {Sov.
  Phys. Solid State}\ }\textbf {\bibinfo {volume} {21}},\ \bibinfo {pages}
  {1401} (\bibinfo {year} {1979})}\BibitemShut {NoStop}%
\bibitem [{\citenamefont {{Kugel'}}\ and\ \citenamefont
  {Khomskii}(1982)}]{Kugel_ufn82}%
  \BibitemOpen
  \bibfield  {author} {\bibinfo {author} {\bibfnamefont {K.~I.}\ \bibnamefont
  {{Kugel'}}}\ and\ \bibinfo {author} {\bibfnamefont {D.~I.}\ \bibnamefont
  {Khomskii}},\ }\href {https://doi.org/10.1070/PU1982v025n04ABEH004537}
  {\bibfield  {journal} {\bibinfo  {journal} {Soviet Physics Uspekhi}\ }\textbf
  {\bibinfo {volume} {25}},\ \bibinfo {pages} {231} (\bibinfo {year}
  {1982})}\BibitemShut {NoStop}%
\bibitem [{CEC(2009)}]{CECAM}%
  \BibitemOpen
  \href@noop {} {\emph {\bibinfo {title} {{CECAM} Workshop. Orbital
  Magnetization in Condensed Matter}}}\ (\bibinfo {address} {CECAM-HQ-EPFL.
  Lausanne. Switzerland},\ \bibinfo {year} {2009})\BibitemShut {NoStop}%
\bibitem [{\citenamefont {Hirst}(1997)}]{Hirst_rmp97}%
  \BibitemOpen
  \bibfield  {author} {\bibinfo {author} {\bibfnamefont {L.~L.}\ \bibnamefont
  {Hirst}},\ }\href {https://doi.org/10.1103/RevModPhys.69.607} {\bibfield
  {journal} {\bibinfo  {journal} {Rev. Mod. Phys.}\ }\textbf {\bibinfo {volume}
  {69}},\ \bibinfo {pages} {607} (\bibinfo {year} {1997})}\BibitemShut
  {NoStop}%
\bibitem [{\citenamefont {Sirotin}\ and\ \citenamefont
  {{Shaskol'skaya}}(1982)}]{Sirotin_Shask_92}%
  \BibitemOpen
  \bibfield  {author} {\bibinfo {author} {\bibfnamefont {Y.~I.}\ \bibnamefont
  {Sirotin}}\ and\ \bibinfo {author} {\bibfnamefont {M.~P.}\ \bibnamefont
  {{Shaskol'skaya}}},\ }\href@noop {} {\emph {\bibinfo {title} {Fundamentals of
  Crystal Physics}}}\ (\bibinfo  {publisher} {Mir Publishers},\ \bibinfo
  {address} {Moscow},\ \bibinfo {year} {1982})\BibitemShut {NoStop}%
\end{thebibliography}

%

\end{document}